\documentclass[10pt, conference]{IEEEtran}
\usepackage{amsmath,amssymb,amsfonts}
\usepackage{epsfig}
\usepackage{graphicx}
\usepackage{epsf}
\usepackage{multirow}
\begin{document}

\title{Tradeoff Between Source and Channel Coding for Erasure Channels}

\author{\authorblockN{Sriram N. Kizhakkemadam, Panos Papamichalis, Mandyam Srinath, Dinesh Rajan}
\authorblockA{Department of Electrical Engineering \\
Southern Methodist University \\
Dallas, TX - 75205, USA \\
Email: \{skizhakk, panos, mds, rajand\}@engr.smu.edu} }

\maketitle

\begin{abstract}
In this paper, we investigate the optimal tradeoff between source
and channel coding for channels with bit or packet erasure. Upper
and Lower bounds on the optimal channel coding rate are computed to
achieve minimal end-to-end distortion. The bounds are calculated
based on a combination of sphere packing, straight line and
expurgated error exponents and also high rate vector quantization
theory. By modeling a packet erasure channel in terms of an
equivalent bit erasure channel, we obtain bounds on the packet size
for a specified limit on the distortion.

\end{abstract}

\section{Introduction}
In~\cite{Shannon48} Shannon presented his celebrated result on the
asymptotic optimality of separable source and channel coding.
However, for finite block length systems, the importance and
superior performance of joint source channel coders has been well
recognized and is an area of active research~(see for example
\cite{Gastpar}). Specific research thrusts have included
investigating source coders that incorporate channel information in
the design, channel coders that provide unequal error protection to
various source bits, and iterative source-channel decoders.

An important issue in joint source channel coding is the tradeoff
between source and channel coding rates. For a fixed source vector
dimension and channel capacity, there is a tradeoff between the
source and channel coding rates. A high rate channel code implies
more bits for the source coder which results in a high quality
representation at the source but has a higher probability of being
received in error. Similarly, a low rate channel code results in
fewer bits for the source coder; consequently, the representation is
of lower quality at the source but there is a higher probability of
being received without error at the receiver. This tradeoff has been
quantified for binary symmetric channels~(BSC)~\cite{Hochwald97} and
Gaussian channels~\cite{Hochwald98}.

This paper addresses the problem of optimal allocation of rate
between a source encoder and a channel encoder for transmission over
erasure channels. The system under investigation is a concatenation
of a vector quantizer with a channel coder and the objective is to
minimize the end-to-end distortion. Upper and lower bounds on the
channel coding rate are constructed that minimizes the end-to-end
distortion. The upper bound on channel coding rate is derived using
the sphere packing and straight line exponents as a bound for
performance of the channel code. Similarly, the lower bound on rate
is derived based on the expurgated error exponent for the erasure
channel. The proposed bounds suggest that the optimal channel coding
rate is substantially smaller than the channel capacity.
Asymptotically, as the erasure probability $\epsilon \rightarrow 0$,
the optimal channel coding rate equals 1. The resulting upper and
lower bounds are then adapted to obtain the optimal coding rate for
packet erasure channels. \par The closed form approximations for the
optimal coding rate are derived under the assumption of
asymptotically small erasure probabilities. Also, a high rate
quantization regime is considered and hence, distortion achieved
asymptotically with large k equals the rate distortion bound. The
proposed bounds are independent of the source distribution for
sources with fixed dimensionality and finite support. \par The rest
of the paper is organized as follows. In Section~2, we define the
system and present the notation and assumptions made in this paper.
In Section~3 we evaluate the upper and lower bounds on the rate for
erasure channels using the expurgated, sphere packing and straight
line bounds. In Section~4 we present some numerical results and
conclude in Section~5.

%


\begin{figure*}[tbph]
    \begin{center}
    \mbox{\epsfbox{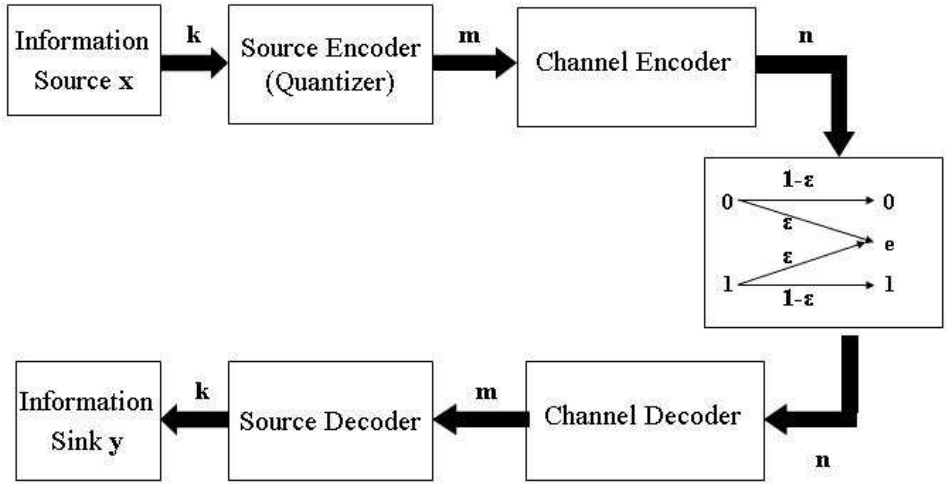}}
  \end{center}

    \label{Fig. 1.}
    \caption{System Block Diagram}
\end{figure*}

A model of the communication system under investigation is given in
Fig. 1. Consider a random vector $\textit{X} \in {\Re}^{k}$ that has
a probability density function \textit{f} over support set $A$, a
closed bounded subset of ${\Re}^{k}$ with nonempty interior. Let X
be quantized by a vector quantizer $Q: {\Re}^{k}\rightarrow C$ where
$C=\left[\textbf{y}_{1}, \textbf{y}_{2}, ...,\textbf{y}_{M}\right]$
is the codebook of the vector quantizer with $m=\log M$ bits per
source symbol. All logarithms are to base 2. Consequently, the
quantizer can be modeled as~\cite{GershoGray},
\begin{equation}
Q\left(\textbf
x\right)=\sum_{i=1}^{M}\textbf{y}_{i}\textbf{1}_{S_{i}}\left(\textbf
x\right)
\end{equation}
where $\lbrace{S_{i}\rbrace}_{i=1}^{M}$ is a partition of
${\Re}^{k}$ into disjoint regions, each of which is represented by
$\textbf{y}_{i}$ and $\textbf{1}_{S_{i}}\left(.\right)$ is the
indicator function which equals 1 if \textbf x lies in the $i^{th}$
cell of the partition. The average distortion using this quantizer
is given by,
\begin{eqnarray}
D_{m}\left(Q\right) &=& \sum_{i=1}^{M}\int_{S_i}||\textbf{x}-\textbf{y}_{i}||^{p}f\left(x\right)dx  \\
&=&2^{-pRr + O\left(1\right)}, \label{eqn:dist_quant}
\end{eqnarray}
where~(\ref{eqn:dist_quant}) follows from Zador's distortion
formula~\cite{Zador82} and $p$ is the power of the distortion
measure. Traditional quantization theory has worked on computing
optimal quantizers that achieve the infimum of
$D_{m}\left(Q\right)$. The quantizer design has been expanded to
include the effect of channel errors; however, the problem is
extremely challenging and little analytical results are known in
such cases. \newline In our system, we consider that the $m$-bit
source codewords are first randomly permuted using a mapping $\pi$
and then passed to a channel encoder of rate $r=m/n$ before
transmission over a binary erasure channel with erasure probability
$\epsilon$. For simplicity, we have included the index assignment
$\pi$ as part of the source encoder. The results are independent of
the index assignment. The channel encoder generates a unique $n$-bit
channel codeword for each of the $m$-bit source codewords. The added
redundancy $n-m$ is used to protect the source code word from
channel impairments. The transmission rate per source component is
$R = n/k$, and the quantization rate is $R_{s}=m/k=Rr$. Following
the notations used in~\cite{Hochwald97}, we denote
$a_{i}=O\left(b_{i}\right)$ if $|a_{i}|/b_{i} \leq c$ for some $c >
0$ and $i$ sufficiently large. We denote
$a_{i}=\Omega\left(b_{i}\right)$ if $|a_{i}|/b_{i} \geq c$ for some
$c>0$ and sufficiently large $i$. Finally,
$a_{i}=o\left(b_{i}\right)$ if $\lim_{i\rightarrow\infty}a_{i}/b_{i}
= 0$.
\section{Binary Erasure Channel}
We now consider obtaining bounds on the coding rate for a binary
erasure channel (BEC). The end-to-end distortion for the system in
Fig.~1, is readily given by~\cite{Hochwald97}
\begin{equation}
D_{R}\left(Q,\epsilon\right) =
\sum_{i,j=1}^{M}q\left(j|i\right)\int_{S_i}||\textbf{x}-\textbf{y}_{i}||^{p}f\left(x\right)dx
\end{equation}
where $\epsilon$ is the bit erasure probability and
$q\left(j|i\right)$ is the conditional probability that the channel
decoder decides in favor of the $j^{th}$ channel codeword when the
$i^{th}$ codeword was transmitted.

\subsection{Lower bound on channel coding rate}
\par
The lower bound on the channel coding rate is obtained by upper
bounding the distortion at the decoder. Assuming small bit erasure
probability and following \cite{Hochwald97}, the total distortion
may be upper bounded as,

\begin{equation}
D_{R}\left(Q,\epsilon \right)
\leq D_{m}\left(Q\right) + O(1)\max_{1\leq i \leq M}P_{e|i}
\label{eqn:total_dist_lb}
\end{equation}
In (\ref{eqn:total_dist_lb}), the total distortion is the sum of the
distortion due to the vector quantizer~( $D_{m}\left(Q\right)$), and
the distortion due to the errors in transmission. The positive O(1)
term is due to the fact that $f$ has support $A$, and $y_{j}$ is
contained in $A$ for all $j$ \cite{Hochwald97}. The problem of
interest is posed as follows: ``Given a binary erasure channel with
$R$ and source $\textit{X} \in {\Re}^{k}$, find the optimal rate $r$
that minimizes the distortion $D_{R}\left(Q,\epsilon,\right)$''

For an arbitrary binary discrete memoryless channel, Shannon's
channel coding theorem guarantees that for channel code rates $r$
below capacity, the probability of error is upper bounded
by~\cite{Gallager68}
\begin{equation}
\max_{1\leq i \leq M}{P}_{e|i} \leq {2}^{-nE_{ex}(r) + o(r)},
\end{equation}
where, $E_{ex}\left(r\right)$ is the expurgated error exponent and
is an exponentially decreasing function of the rate. The dependence
of $P_{e|i}$ on $n$ indicates that the decoding error probability
can be decreased by increasing the length of the channel codewords.
The expurgated error exponent is given by \cite{Gallager68},
\begin{equation}
{E}_{ex}\left(r\right) = \sup_{\rho \geq 1}\left[-\rho r +
\max_{\textbf{q}} {E}_{x}\left(\rho, \textbf{q}\right)\right]
\end{equation}
where,
\begin{gather}
{E}_{x}\left(\rho, \textbf{q} \right) = \nonumber
\\ -\rho\log\sum_{k=0}^{K-1}\sum_{i=0}^{K-1}q\left(k\right)q\left(i\right)\left[\sum_{j=0}^{J-1}\sqrt{P\left(j|k\right)P\left(j|i\right)}\right]^{1/\rho}.
\label{eqn1}
\end{gather}
Note that ${\mathbf q} = [q(0)\: q(1)\: \ldots q(K-1)] $ represents
the probability of the input channel alphabets and $P(j|i)$ is the
probability of receiving output symbol $j$ when input symbol $i$ is
transmitted. In $\left(\ref{eqn1}\right)$, $K$ and $J$ represent,
respectively, the cardinality of the input and output alphabets of
the channel. For a binary erasure channel, $J=3$ and $K=2$. For the
binary erasure channel,the transition probability matrix is given by

\begin{table}[h]
    \begin{center}
        \begin{tabular}{|c c|c|c|c|} \hline
        & & \multicolumn{3}{|c|}{Output, j}\\
        \hline
        \multirow{3}{*}
        & &1 & e & 0\\ \hline
        {Input, i/k} & 1 & $1-\epsilon$ & $\epsilon$ & 0\\
        & 0 & 0 & $\epsilon$ & $1-\epsilon$ \\
        \hline
        \end{tabular}
        \end{center}
    \caption{Transition Probability Matrix with elements $P(j|i)$ for BEC.}
\end{table}


For a symmetric channel, the \textbf{q} that maximizes the error
exponent is the uniform probability assignment~\cite{Jelinek68}.
Thus for the binary erasure channel, $\textbf{q}
=\left[q\left(0\right)\: q\left(1\right)\right]=\left[0.5\:
0.5\right]$. Substituting the upper bound for the probability of
error into~(\ref{eqn:total_dist_lb}), we obtain the end-to-end
distortion as
\begin{equation}
D_{R}\left(Q,\epsilon,\pi\right) \leq 2^{-pRr + O\left(1\right)}
+{2}^{-kRE_{ex}(r) + o(r)} \label{eqn:dist_lb2}
\end{equation}
Consider the case of large $R$: To ensure that neither of the two
terms on the right hand side of~(\ref{eqn:dist_lb2}) dominates the
distortion upper bound, we choose the exponents of the two terms to
be within o(1) of each other \cite{Hochwald97},
\cite{ZegerManzella}. Hence, we set
\begin{equation}
E_{ex}\left(r\right) = \frac{p}{k}r_{ex} + o\left(1\right),
\label{eqn:linear_reln_lb}
\end{equation}
to obtain the channel coding rate that optimizes the end-to-end
distortion at the decoder. This optimal rate is characterized by
Theorem~1, which is similar to Theorem~1 in~\cite{Hochwald97}.


\textit{Theorem 1:} The upper bound on the minimum $p^{th}$ power
distortion, averaged over all index assignments of a $k$-dimensional
cascaded good vector-quantizer and channel encoder that transmits
over a binary erasure channel with bit erasure probability
$\epsilon$, is achieved with a channel code rate $r_{ex}$ satisfying

\begin{gather}
r_{ex} = 1 -
2^{-c_\epsilon}\left(\frac{\log\log\left(1/\epsilon\right) + \log e
+ c_\epsilon}{\log\left(1/\epsilon\right)}\right) \nonumber
\\+ O\left(\frac{\log\log\left(1/\epsilon\right)}{\log^{2}\left(1/\epsilon\right)}\right) + o\left(1\right),
\label{eqn:optimal_rate_lb}
\end{gather}
where, $c_\epsilon$ satisfies
\begin{equation}
\frac{p}{k}2^{c_\epsilon} -
\frac{\left(p/k\right)\left(\log\log\left(1/\epsilon\right)+\log e +
c_{\epsilon}\right) -2^{-c_\epsilon}}{log\left(1/\epsilon\right)} -
1 = 0.
\end{equation}

\textit{Proof:}
\par
For a BEC, evaluating (8), we obtain
\begin{equation}
\max_\textbf{q} {E}_{x}\left(\rho, \textbf{q}\right) = \rho\left[1 -
\log\left(1 + \epsilon ^{1/\rho}\right)\right]
\end{equation}
and thus the expurgated error exponent becomes
\begin{equation}
E_{ex}\left(r\right) = \sup_{\rho \geq 1}\left\{ \rho\left[1-r-
\log\left(1 + {\epsilon}^{1/\rho}\right)\right]\right\}
\label{eqn:expurgated_exponent}
\end{equation}
The $\rho$ which maximizes the error exponent and also
satisfies~(\ref{eqn:linear_reln_lb}) is given by
\begin{equation}
\rho =
\frac{\log\left(1/\epsilon\right)}{\log\log\left(1/\epsilon\right) +
c_\epsilon}
\end{equation}
Substituting for $\rho$ and $c_{\epsilon}$
into~(\ref{eqn:linear_reln_lb}) we obtain the optimal
rate~(\ref{eqn:optimal_rate_lb}) and hence the theorem is proved.
\hfill $\square$

Note that the expression for the expurgated error joint source
channel rate is similar to the BSC case \cite{Hochwald97} with the
difference being the argument of the $\log$ term. Appendix~1
in~\cite{Hochwald97} provides details on the derivation for $\rho$
and $c_{\epsilon}$. A further simplification in the expression for
the rate can be obtained by neglecting the $O(1)$ and $o(r)$ terms
and  equating ~(\ref{eqn:expurgated_exponent}) to the exponent of
the source coding distortion yielding
\begin{equation}
r_{ex}=\frac{\rho}{\frac{p}{k} + \rho}\left[1 - \log\left(1 +
\epsilon^{1/\rho}\right)\right]
\end{equation}
Numerical values of $r_{ex}$ is given in Figure~2 and are explained
in Section IV.

\subsection{Upper bound on channel coding rate}
Following the analysis in~\cite{Hochwald97}, the upper bound on the
average distortion minimized over all channel code rates for large
$R$ and small bit error probability for the binary erasure channel
can be obtained as
\begin{gather}
D_{R}\left(Q,\epsilon,\pi\right)\geq
D_{m}\left(Q\right)\left(1-P_{e}\right) +
\Omega\left(1\right)\frac{1}{M}\sum_{k=1}^{M}P_{e|k} \nonumber
\\=2^{-pRr + O\left(1\right)}\left(1-P_{e}\right) + \Omega\left(1\right)P_{e}
\end{gather}
where $P_e$ is the probability of error occurring in the channel.  A
lower bound on this probability of error is given by,

\begin{equation}
{P}_{e} \geq {2}^{-nE_{sl}(r) + o(n)}  = 2^{-kRE_{sl}\left(r\right)
+ o\left(R\right)}
\end{equation}
where $E_{sl}$ is the straight line exponent. The straight line
exponent $E_{sl}\left(r\right)$ is a linear function of $r$ which is
tangent to the sphere packing exponent $E_{sp}\left(r\right)$ and
also satisfies $E_{sl}\left(0\right) = E_{ex}\left(0\right)$. The
sphere packing exponent \cite{Gallager68} is given by,
\begin{equation}
{E}_{sp}\left(r\right) = \sup_{\rho \geq 0}\left[-\rho r +
\max_{\textbf{q}} {E}_{o}\left(\rho, \textbf{q}\right)\right]
\end{equation}
and,
\begin{equation}
\max_\textbf{q} {E}_{o}\left(\rho, \textbf{q}\right) =
-\log\sum_{j=0}^{J}\left[\sum_{k=0}^{K}q(k)P(j|k)^{1/\left(1+\rho\right)}\right]^{1+\rho}
\end{equation}
The straight line exponent can be written as,
\begin{equation}
E_{sl}\left(r_{sl}\right) = E_{ex}\left(0\right) +
r_{sl}\frac{\left[E_{sp}\left(r'\right) -
E_{ex}\left(0\right)\right]}{r'} \label{eqn:Esl}
\end{equation}
where, $r'$ is the rate at which the straight line exponent meets
the sphere packing exponent tangentially. The straight line exponent
has also been characterized in~\cite{McEliece} for a binary erasure
channel. Now, the end-to-end distortion is thus bounded as
\begin{equation}
D_{R}\left(Q,\epsilon,\pi\right)\geq 2^{-pRr + O(1)} +
2^{-kRE_{sl}(r) + o(R)} \label{eqn:total_dist_ub1}
\end{equation}

The channel coding rate that minimizes this bound is now
characterized in Theorem~2.

\textit{Theorem 2:} An upper bound on the channel code rate $r$ that
minimizes the $p^{th}$ power distortion averaged over all random
index assignments of a $k$-dimensional cascaded good vector
quantizer for a binary erasure channel with small effective bit
erasure probability $\epsilon$ and large $R$ is given by
\begin{equation}
r_{sl} = \frac{E_{ex}\left(0\right)}{\frac{p}{k} -
\frac{E_{sp}\left(r'\right) - E_{ex}\left(0\right)}{r'}}
\label{eqn:optimal_rate_ub}
\end{equation}

\textit{Proof:} As in the earlier case, for large $R$, to prevent
either of the terms in the distortion
bound~(\ref{eqn:total_dist_ub1}) from dominating the other, we set
the straight line exponent to be linearly proportional to the
exponent term of the noiseless-optimal distortion within
$o\left(1\right)$ of each other. Thus,
\begin{equation}
E_{sl}\left(r\right) = \frac{p}{k}r_{sl} + o\left(1\right)
\label{eqn:linear_reln_ub2}
\end{equation}
Substituting ~(\ref{eqn:Esl}) in ~(\ref{eqn:linear_reln_ub2}), the
theorem is proved \hfill $\square$.

Note that to completely characterize $r_{sl}$ we need to explicitly
evaluate the sphere packing exponent $E_{sp}(r)$. It is easily seen
that a uniform probability assignment for the input states to the
channel $q\left(.\right)$ maximizes $E_{o}\left(\rho,
\textbf{q}\right)$ and thus $E_{sp}(r)$ can be evaluated as,
\begin{equation}
E_{sp}\left(r\right) =
\sup_{\rho\geq0}\left\{\rho\left(1-r\right)-\log\left[\left(1-\epsilon\right)+\epsilon2^{\rho}\right]\right\}
\label{eqn:Esp}
\end{equation}

Note that $(\ref{eqn:Esp})$ is a concave function of $\rho$ and
hence the supremum can be replaced by the max operator. The $\rho$
which maximizes ~(\ref{eqn:Esp}) satisfies
\begin{equation}
r=\frac{\left(1-\epsilon\right)}{\left(1-\epsilon\right) +
2^{\rho}\epsilon}
\end{equation}
We can use this relation between the rate and $\rho$ to express the
sphere packing exponent in terms of the channel encoding rate for a
given erasure channel as,
\begin{equation}
E_{sp}\left(r\right) = r\log r + ~(1-r)\log(1-r)
-r\log\left(\frac{1-\epsilon}{\epsilon}\right) -\log\epsilon
\label{eqn:Esp_r}
\end{equation}
At $r'$, the slope of the sphere packing exponent equals the slope
of the straight line exponent. Thus,
\begin{equation}
\frac{\partial E_{sp}}{\partial
r}\mid_{r=r'}=\frac{E_{sp}\left(r'\right)-E_{ex}\left(0\right) }{r'}
\label{slope}
\end{equation}
Differentiating $(\ref{eqn:Esp_r})$ and substituting in
~(\ref{slope}), we get
\begin{equation}
r'=1 - 2^{E_{ex}~(0)-\log\left(1/\epsilon\right)}
\label{eqn:r_prime}
\end{equation}
It turns out that $E_{sp}(r')$ is nearly 0 for small values of
$\epsilon$.

\section{Numerical Results}
 The bounds derived above for the erasure channel can be easily extended to the case of packet
erasures. We use a simplified model for the packet erasure channel
and assume that a packet erasure occurs if any of the bits within
the packet suffers an erasure. Although this assumption simplifies
the packet erasure channel model, it is useful in obtaining closed
form bounds on the coding rate over such channels. For a packet of
size~$P$ bits, the probability of a packet erasure $\delta$ is given
by $\delta = 1 - \left(1 - \epsilon\right)^{P}$, where as before
$\epsilon$ denotes the probability of bit erasure. The error
exponent for a binary erasure channel with erasure probability
$\epsilon$ and a $2^{P}$-ary erasure channel with erasure
probability $\delta$ is the same. Hence, given the packet erasure
probability $\delta$, we consider an equivalent binary erasure
channel with bit erasure probability $\epsilon = 1 - \left(1 -
\delta\right)^{1/P}$ and find the bounds on the rate and distortion
for the corresponding BEC.

The plot of the upper and lower bound on channel coding rate as a
function of the erasure channel probability for k=4 and squared
distortion measure is given in Fig.~2. The bounds for various packet
sizes $P=1$, 10 and 100 are shown in Fig.~2 . It is observed that
for a given packet size, as the erasure probability increases, the
channel coding rate decreases indicating that more bits need to be
invested on channel coding to combat a hostile channel. Further, for
a given packet erasure probability, as the packet size increases,
the channel coding rate increases implying that more bits can be
allocated for source coding with larger packet size. Fig.~3 offers a
different perspective on the results. From the bounds on the channel
coding rate, we can get the bounds on the distortion due to channel
coding. By virtue of our optimal joint source-channel coding
criterion, the total distortion will be twice the distortion due to
channel coding. Hence, given an end-to-end limit on the distortion,
we can get the minimum packet length to be chosen from Fig.~3. The
squared distortion metric with $k=4$, $R=10$ and packet erasure
probability $\delta=10^{-3}$ was chosen. The $o\left(r\right)$ and
$o\left(R\right)$ terms were neglected in the terms for distortion
due to noisy channel decoding in $(\ref{eqn:dist_lb2})$ and
$(\ref{eqn:total_dist_ub1})$.The asymptotic nature of the curve
indicates that large packet size is not required for packet erasure
channels with small erasure probabilities.

%
%

\begin{figure}[tbph]
    \begin{center}
    \epsfxsize = 3.5in
    \epsfbox{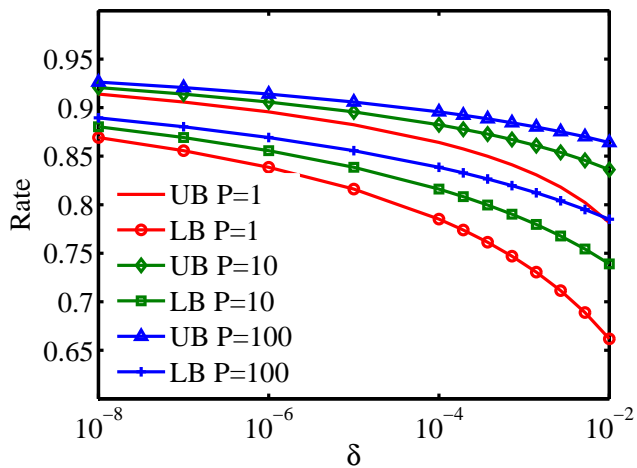}
  \end{center}

    \label{Fig. 2.}
    \caption{The upper (UB) and lower bounds (LB) on the optimal channel
    coding rate are plotted for various values of packet size P.}
\end{figure}

\begin{figure}[h]

\begin{center}
 \epsfxsize = 3.5 in \mbox{
\epsfbox{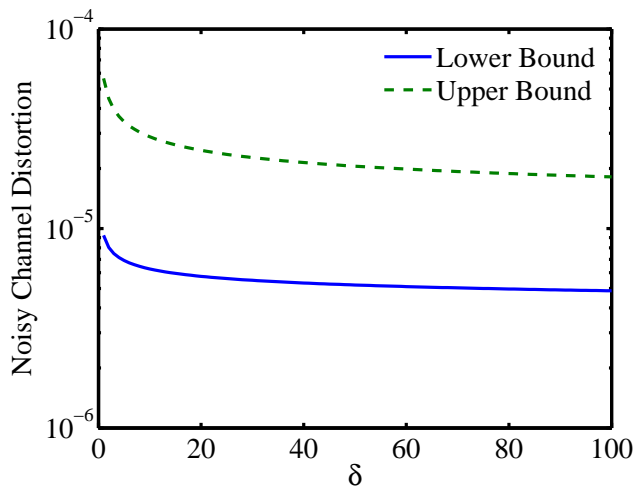}}
\end{center}
\label{Fig. 3.} \caption{The Distortion for various packet lengths
for a packet erasure channel with packet erasure probability $\delta
= 10^{-3}$ and R = 10.}
\end{figure}

\section{Conclusion}
The results presented in this paper provide a mechanism for optimal
concatenation of source and channel coders. Analytic results are
provided for lower and upper bounds for a binary erasure channel and
for packet erasure channels. The results on packet erasure channel
enable us to obtain the bounds on the packet size for a specified
bound on the distortion and given packet erasure probability.
Alternately, for a given packet erasure probability, we can find
bounds on the channel encoding rate for various packet lengths. By
studying the optimal rate allocation for a bit and packet erasure
channel, one can apply these results for transmission in a wide
range of scenarios, including wireline channels with congestion.  In
future work, these bounds should be expanded to include transmission
over more sophisticated channel models.


\section*{Acknowledgment}
This work has been supported in part by Nokia Inc.



\begin{thebibliography}{1}
\bibitem{Shannon48}
C. E. Shannon, ``A mathematical theory of communication,''
\emph{Bell Syst.\ Tech.\ J.}, vol.\ 27, pt.~I, pp.~379--423, 1948;
     pt.~II, pp.~623--656, 1948.
\bibitem{Hochwald97}
B. Hochwald, K. Zeger, ``Tradeoff between source and channel
coding,'' \emph{IEEE Trans. Inform. Theory}, vol.\ IT-43,
pp.~1412--1424, Sept. 1997.
\bibitem{Hochwald98}
B. Hochwald, ``Tradeoff between source and channel coding on a
Gaussian channel,'' \emph{IEEE Trans. Inform. Theory}, vol.\ IT-44,
No.~ 7, pp.~1412--1424, Nov. 1998.
\bibitem{GershoGray}
A.~ Gersho and R.~ M.~ Gray, \emph{Vector Quantization and Signal
Compression}. Boston, MA Kluwer Academic, 1992.
\bibitem{Zador82}
P. Zador, ``Asymptotic quantization error of continuous signals and
the quantization dimension,'' \emph{IEEE Trans. Inform. Theory},
vol.\ IT-28, pp.~ 139-149, Mar. 1982.
\bibitem{Gallager68}
R. Gallager, \emph{Information Theory and Reliable Communication}.
New York: Wiley, 1968.
\bibitem{Jelinek68}
F. Jelinek, ``Evaluation of expurgated bound exponents,'' \emph{IEEE
Trans. Inform. Theory}, vol.\ IT-14, pp.~ 501-505, May. 1968.
\bibitem{ZegerManzella}
K. Zeger and V. Manzella, ``Asymptotic bounds on optimal noisy
channel quantization via random coding'', \emph{IEEE Trans. Inform.
Theory}, vol.\ IT-40, pp. 1926-1938, Nov. 1994.

\bibitem{Gastpar}
M. Gastpar, B. Rimoldi, and M. Vetterli, ``To code, or not to code:
lossy source-channel communication revisited'' \emph{IEEE Trans.
Inform. Theory}, IT-49 , pp. 1147-1158, May 2003.

\bibitem{McEliece}
R. McEliece and J. Omura, ``An improved upper bound on the block
coding error exponent for binary-input discrete memoryless
channels'' \emph{IEEE Trans. Inform. Theory}, IT-23, Issue 5, pp.
611 - 613, Sep. 1977.

\end{thebibliography}
\end{document}